\documentclass[pre,twocolumn,showpacs,preprintnumbers,amsmath,amssymb,floatfix]
{revtex4}

\usepackage{graphicx}
\usepackage{color}
\usepackage{verbatim}
\usepackage{float}
\newcommand \be {\begin{equation}}
\newcommand \ee {\end{equation}}
\newcommand \ben {\begin{eqnarray}}
\newcommand \een {\end{eqnarray}}

\newcommand \liqr {\rho_{\ell}}

\begin{document}

\title{
Phase field crystal modeling as a unified atomistic approach to defect dynamics
}

\author{Joel Berry$^{1,3,*}$, 
Nikolas Provatas$^{2}$, 
J\"org Rottler$^3$, and 
Chad W.\ Sinclair$^4$
}

\affiliation{$^1$Department of Materials Science and Engineering,
McMaster University, 1280 Main Street West, Hamilton, Ontario, L8S 4L7, Canada}
\affiliation{$^2$Physics Department, McGill University,
3600 rue University, Montr\'eal, Qu\'ebec, H3A 2T8, Canada}
\affiliation{$^3$
Department of Physics and Astronomy, The University of British Columbia, 
6224 Agricultural Road, Vancouver, British Columbia, V6T 1Z1, Canada}
\affiliation{$^4$
Department of Materials Engineering, The University of British Columbia, 
309-6350 Stores Road, Vancouver, British Columbia, 
V6T 1Z4, Canada}
\altaffiliation{
{\it Corresponding author}:
jmberry@princeton.edu
{\it Current address}:
Department of Mechanical and Aerospace Engineering, Princeton University, 
Princeton, NJ 08544, USA}

\date{\today}

\begin{abstract}
Material properties controlled by evolving defect structures,
such as mechanical response,
often involve processes spanning many length and time scales
which cannot be modeled using a single approach.
We present a variety of new results that demonstrate the ability of
phase field crystal (PFC) models to describe complex defect evolution
phenomena on atomistic length scales and over long, diffusive time scales.
Primary emphasis is given to the unification of
conservative and nonconservative dislocation creation mechanisms
in three-dimensional FCC and BCC materials.
These include Frank-Read-type glide mechanisms involving closed dislocation
loops or grain boundaries as well as
Bardeen-Herring-type climb mechanisms involving precipitates, inclusions, 
and/or voids.
Both source classes are naturally and simultaneously captured at the
atomistic level by PFC descriptions, with arbitrarily complex defect
configurations, types, and environments.
An unexpected dipole-to-quadrupole source transformation is identified,
as well as various new and complex geometrical features of loop nucleation
via climb from spherical particles.
Results for the strain required to nucleate a dislocation loop from such a 
particle are in agreement with analytic continuum theories.
Other basic features of FCC and BCC dislocation structure and dynamics are
also outlined, and initial results for dislocation-stacking fault tetrahedron
interactions are presented.
These findings together 
highlight various capabilities of the PFC approach
as a coarse-grained atomistic
tool for the study of three-dimensional crystal plasticity.
\end{abstract}

\maketitle

\section{Introduction}
\label{sec:intro}

The macroscopic mechanical properties of crystals and polycrystals are
primarily
consequences of complex collective interactions between atomic level defects.
The characteristic scales of these interaction processes can span many
orders of magnitude in length and time, presenting major fundamental
challenges to the development of a unified modeling approach.
In particular, line defects or dislocations, which
are the central mediators of plasticity in many systems,
can evolve rapidly via conservative
mechanisms (e.g., glide and cross-slip) or
relatively slowly via nonconservative mechanisms
mediated by interactions with point defects (e.g., climb).
Characteristic length scales of dislocation structures range from
atomic dimensions to micron-level and up for collective, organized arrays.
A complete model of dislocation dynamics would therefore ideally 
include the fundamental physics of dislocation 
creation, interaction, annihilation/absorption, etc,
via both conservative and nonconservative mechanisms, 
accessible across all relevant length and time scales.
This is not feasible with any presently available model. 
Other classes of defects (point, planar, and bulk) 
should also be considered in a more general
model of crystalline materials subjected to driving forces.

Elements of conservative dislocation processes 
are often quite readily modeled at the atomic level using approaches
such as molecular dynamics (MD) 
\cite{bulatov2006computer,MDdislocGB2002}.
These conservative mechanisms have also been built into larger length
scale mesoscopic modeling approaches such as discrete dislocation
dynamics (DDD) \cite{DDintro1990,DDreview2009,bulatov2006computer} and 
continuum phase field (PF) dislocation models 
\cite{PFdislocs2003,PFDDreview2010,bulatov2006computer}.
Basic conservative processes are generally implemented in DDD with detailed
rule-based formulations that consider some number of the innumerable possible
defect interaction scenarios. PF models avoid this complexity by
treating select 
dislocation lines as interfaces in a continuum field description.
These interfaces interact and evolve automatically in response to local 
driving forces.
The cost is greater computational demand, since the PF equations
must be solved throughout the entire system, not only at localized
dislocation positions.

Nonconservative dislocation processes 
present another set
of challenges to plasticity models, as these motions are mediated by vacancy
diffusion and are therefore inherently difficult to access on conventional
atomistic simulation time scales.
Climb becomes relevant and often dominant at high temperatures or large
vacancy concentrations, and is fundamental to such phenomena 
as creep, annealing, recrystallization, and irradiation damage.
Elements of climb have been built into DDD 
\cite{DDclimb1998long,DDclimb2008,DDclimbBH2011,DDclimb2013}
and PF 
\cite{PFclimb2012,PFclimb14}
models, though this issue is still in many ways under development.
Meaningful coarse-grained input parameters and their values, for example,
which would ideally be extracted from microscopic simulations, are lacking.

No atomistic modeling approach has yet proven
capable of consistently describing both conservative and nonconservative
dislocation processes over both nano and mesoscales.
In this work, we present results for a method that unifies
both types of dislocation motion on atomistic length scales.
Phase field crystal (PFC) models \cite{pfc02,pfc04,pfcdft07}
describe diffusive dynamics in 
condensed matter systems with atomistic resolution,
and are therefore potentially capable
of bridging the gap between fast glide plasticity and slow climb plasticity
at the nanoscale.
Defect superstructures of arbitrary complexity can be studied, including
polycrystals with nearly any variation/combination of dislocation, 
grain boundary, precipitate, and stacking fault configurations, for example.
The ability to naturally describe arbitrary defect structures is a 
feature of atomistic approaches
that is inherently absent from mesoscale approaches.
The larger length scales described by DDD and
PF models cannot currently be reached by PFC, though
one may imagine using PFC simulations to generate input parameters for such
models or numerically coupling PFC with DDD or PF. 
Coarse-grained complex amplitude PFC models also provide an interesting means
of self-consistently reaching larger length scales
\cite{pfcRGnigel06a,pfcadmesh07,
pfcRGcoexist10,pfcbinaryamp10,pfcRGkarma10,nanaamp13}.

The goals of the present work are to demonstrate that simple PFC models
naturally capture well-established conservative dislocation creation
mechanisms as well as new elements of the relatively poorly understood
nonconservative dislocation creation mechanisms, which cannot be easily
studied with other methods.
These goals are part of a larger effort to exploit the novel features of
the PFC approach within traditional areas of materials science, including 
crystal plasticity, 
structural phase transformations, and microstructure evolution
\cite{nikpfcstruct10,nikpfcstruct11,xpfcbin11,pfcdefects12,
xpfcternary13,nanaamp13}.
Some initial groundwork covering fundamental dislocation properties in 
FCC materials was reported by the present authors in 
Ref.\ \cite{pfcdefects12}.
Perfect dislocations and simple
grain boundaries in 2D triangular and 3D BCC crystals
have also been examined in various contexts
\cite{pfc04,pfcdisloc06,
mpfc,mpfc09,pfcdeformscheme09,pfcavalanches10,pfcdft07,pfcbcc09,
pfccurved10,pfcfccGB10,pfcpremelt08,pfcpremeltkarma08,pfchotgb11,pfcvapor13}.

The rest of this paper is organized as follows.
In Section \ref{sec:model}, the basic model equations, numerical
solution methods, and strain application procedures are outlined.
In Section \ref{sec:fccbcc}, 
some qualitative and quantitative features of the specific dislocations
relevant to FCC and BCC plasticity are surveyed in the PFC framework.
In Section \ref{sec:frsources}, conservative Frank-Read-type dislocation
sources in FCC materials are studied in two contexts. 
The first considers controlled
nucleation of dislocation loop dipoles, and a new mechanism whereby a
dipole source transforms into a quadrupole source is reported.
The second context considers uncontrolled nucleation of partial
and perfect dislocations from grain boundaries in nanopolycrystalline samples.
In Section \ref{sec:bhsources}, nonconservative Bardeen-Herring-type 
dislocation sources in BCC materials are studied.
The case of uniform loop nucleation from spherical inclusions or 
precipitates is considered. 
A range of complex nucleation
behaviors caused by non-trivial interactions between interface structure,
strain orientation, and dislocation energetics are examined.
Selected results are compared with earlier analytic predictions and
shown to agree well when the analytic theories are adapted to the
scenario considered in our simulations.
Finally, a brief presentation of stacking fault tetrahedron (SFT) formation
and SFT-dislocation interactions is provided in Section
\ref{sec:dislocSFT}.

\section{Model and Methods}
\label{sec:model}

The standard PFC free energy functional, modified to stabilize FCC
stacking faults as described in Ref.\ \cite{pfcdefects12},
is used for all FCC systems studied in the following sections.
It is written
\begin{eqnarray}
\tilde{F}
&=&\int d\vec{r}~\left[
\frac{1}{2}n^2(\vec{r})
-\frac{w}
{6}
n^3(\vec{r}) + 
\frac{u}{12}
n^4(\vec{r}) \right]-\nonumber \\
& &\frac{1}{2} \int\int d\vec{r}~d\vec{r}_2~n(\vec{r})
C_2(|\vec{r}-\vec{r}_2|) n(\vec{r}_2).
\label{pfcfree}
\end{eqnarray}
where $\tilde{F}=F/(k_B T \liqr)$, 
$\liqr$ is a constant reference density,
$n(\vec{r})=\rho(\vec{r})/\liqr-1$ is the rescaled atomic density field,
$\rho(\vec{r})$ is the unscaled atomic number density field,
$w$ and $u$ are coefficients treated as free parameters to
provide additional model flexibility,
and $C_2(|\vec{r}-\vec{r}_2|)$
is the two-point direct correlation function of the fluid, assumed isotropic.
The modified standard PFC kernel
(which approximates the full $C_2(|\vec{r}-\vec{r}_2|)$)
after Fourier transformation reads
\be
\hat{C}_2(k) = -r + 1 - B^x (1-\tilde{k}^2)^2 -
H_0 e^{-(k-k_0)^2/(2\alpha_0^2)},
\label{pfcmod}
\ee
where $r$ is a constant proportional to temperature,
$B^x$ is a constant proportional to the solid-phase elastic moduli,
$\tilde{k}=k/(2\pi)$ is the normalized wavenumber,
$H_0$ is a constant related to stacking fault energy $\gamma_{\rm ISF}$,
$k_0=2\pi\sqrt{41/12}/a$, 
$a$ is the equilibrium lattice constant, 
and $\alpha_0$ is an additional constant related to $\gamma_{\rm ISF}$. 
$n(\vec{r})$ will be allowed
to assume nonzero average values $n_0$,
and it will be implied that $w=0$ and $u=3$ for all FCC simulations.

The structural or XPFC free energy functional class
\cite{nikpfcstruct10,nikpfcstruct11,xpfcbin11}
is used for all BCC systems studied in the following sections.
It is written as Eq.\ (\ref{pfcfree}) with
the Fourier transformed kernel
\be
\hat{C}_2(k)_i = 
e^{-(k-k_i)^2/(2\alpha_i^2)}
e^{-\sigma^2 k_i^2/(2\rho_i \beta_i)}
\label{xpfckernel}
\ee
where $i$ denotes a family of lattice planes at wavenumber $k_i$,
and $\sigma$ is a temperature parameter.
The constants $\alpha_i$, $\rho_i$, and $\beta_i$ are the
Gaussian width (which sets the elastic constants), planar atomic density,
and number of planes, respectively,  
associated with the $i$th family 
{\color{black}
of lattice planes.
}
The envelope of all selected Gaussians $i$ 
composes the final $\hat{C}_2(k)$.
Only one reflection at $k_1=2\sqrt{3}\pi$ will be used here, 
as this is all that is necessary to produce equilibrium BCC structures
with $a \simeq \sqrt{2/3}$.

Two dynamic equations for $n(\vec{r})$ will be considered.
The first is a purely diffusive Model B form,
\be
\frac{\partial n(\vec{r})}{\partial t} = \nabla^2\frac{\delta 
\tilde{F}}
{\delta n(\vec{r})}
\label{eq:pfcdyn1}
\ee
{\color{black}
where $t$ is dimensionless time.
The second equation of motion introduces a faster inertial, quasi-phonon 
dynamic component in addition to diffusive dynamics \cite{mpfc}, 
\be
\frac{\partial^2 n(\vec{r})}{\partial t^2}+\beta\frac{\partial n(\vec{r})}
{\partial t}= \alpha^2 \nabla^2 \frac{\delta 
\tilde{F}}
{\delta n(\vec{r})}
\label{pfcinertia}
\ee
where $\alpha$ and $\beta$ are constants
related to sound speed and damping rate, respectively.
All simulations were performed in 3D using 
a pseudo-spectral algorithm with semi-implicit time stepping
and periodic boundary conditions.
Deformation was applied via strain-controlled methods for both
simple shear and uniaxial tension or compression simulations.
Constant strain rates were employed for both deformation types.
}

\section{Some Fundamental Elements of FCC \& BCC Crystal Plasticity}
\label{sec:fccbcc}

Many aspects of the distinct plastic response of FCC and BCC crystals
can be understood in terms of structural and dynamic differences
between the dominant carriers of plasticity 
in either lattice. 
The primary dislocation type in FCC materials is that with total Burgers vector
$a/2\langle110\rangle$, dissociated into two 
$a/6\langle112\rangle$ Shockley partials connected by a stacking fault.
Such dissociated dislocations provide
FCC crystals with 12 active primary slip systems, all of type 
$\{111\}\langle110\rangle$. 
The very small Peierls stress of the Shockley partials can explain the
low yield stress of FCC materials, and the subsequent formation of large
numbers of dislocation junctions and stacking faults
can explain their excellent work-hardening properties, ductility, 
and the commonly observed formation of mesoscale dislocation patterns
\cite{kubinfccbcc98}.

The primary dislocation type in BCC materials is that with Burgers vector
$a/2\langle111\rangle$.
This dislocation is glissile within
48 potential primary slip systems of type 
$\{110\}\langle111\rangle$,
$\{112\}\langle111\rangle$, or
$\{123\}\langle111\rangle$.
The non-planar core structure of the $a/2\langle111\rangle$
screw dislocation in particular at low temperatures leads to a
much smaller glide mobility than that in edge orientation. The
$a/2\langle111\rangle$ screw dislocation therefore
controls plastic flow in BCC crystals at $T \lesssim 0.15T_{melt}$.
Its high Peierls stress can explain the large yield and flow stresses
of BCC crystals, as well as the absence of mesoscale dislocation patterning
at low to moderate stresses
\cite{kubinfccbcc98}.

\begin{figure}[btp]
 \centering{
   \includegraphics*[width=0.48\textwidth,trim=0 0 0 0]{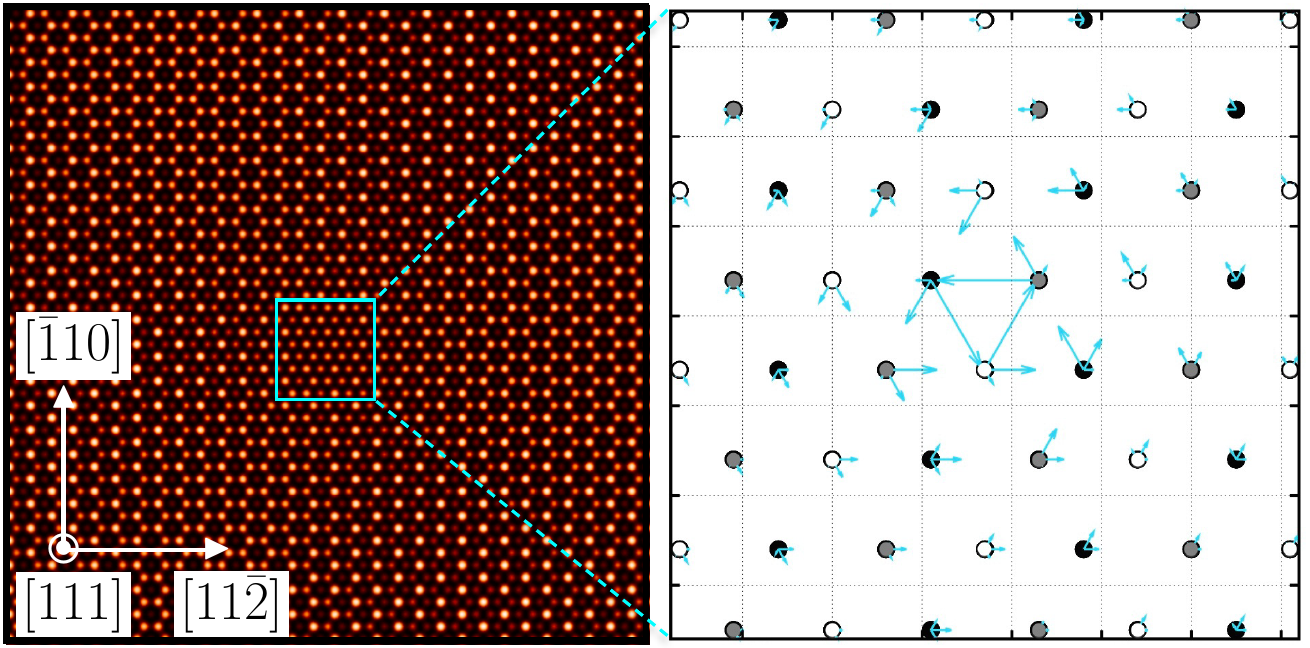}
 }
 \vspace{-.125cm}
\caption[]
{\label{BCCscrewcore}
(Color online)
The BCC $a/2\langle111\rangle$ screw dislocation
from a single-peaked XPFC model.
A cross-section of $n(\vec{r})$ is shown on the left, and 
the differential displacement map
\cite{vitekbcccore1970}
around the core is shown on the right.
Results were generated using
Eqs.\ (\ref{pfcfree}), (\ref{xpfckernel}), and (\ref{eq:pfcdyn1})
with model parameters
$w=1.4$, $u=1$, $n_0=0.05$, $\alpha_1=0.25$, $\sigma=0.12$,
$\rho_1=1$, and $\beta_1=8$.
Nearly identical results are obtained at $\alpha_1=1$ and $\alpha_1=2$.
}
\end{figure}

It can therefore be argued that these dislocation types must be stabilized
and their basic core features reproduced if one wishes to
perform atomistic PFC simulations of FCC and BCC plasticity.
We have found that it is indeed possible to stabilize
both of these dislocation types
within a given PFC model. 
Dissociated $a/2\langle110\rangle$
FCC dislocations were previously studied by the current authors
\cite{pfcdefects12}.
With proper selection of the model parameters,
static properties (dissociation width, Peierls strains, etc)
and glide dynamics of these dislocations were found to be in good agreement
with other atomistic calculations and continuum elastic theories.
For the primary BCC $a/2\langle111\rangle$ screw dislocation,
we have found that the PFC model used in this study
reproduces the same nondegenerate nonpolarized 
core configuration obtained from density functional theory calculations
and MD simulations employing various empirical potentials 
\cite{footnote1} (see Fig.\ \ref{BCCscrewcore}).
The central core features of the primary dislocation types in
both FCC and BCC crystals can therefore be well-captured by PFC models.
This level of accuracy in terms of core structure should not always be expected,
especially for more complex, directionally bonded materials such as
diamond cubic Si or Ge crystals.
Nonetheless, when sufficient accuracy is achieved, 
we find as a general consequence that
the correct slip systems naturally emerge during
simulations of plastic flow, and that
atomistically detailed plasticity mechanisms
also often follow. 

For example, we have confirmed that the glide mobility of 
the BCC $a/2\langle111\rangle$ dislocation 
in screw orientation is significantly lower than that in edge orientation,
as expected based on the non-planar screw core structure.
This mobility difference was inferred from zero-strain simulations of 
glide-mediated dipole
annihilation, in which it was observed that screw dislocation dipoles take
roughly one to two orders of magnitude longer to annihilate than edge dipoles
at equal initial separations.

These same basic structural and energetic features also 
influence climb processes.
For example, compressive and 
tensile strains along the axis of the
dislocation Burgers vector induce the correct positive
and negative climb directions in PFC simulations.
Jogged dislocations subjected to the same type of strain 
climb in the expected fashion
in which jogs diffusively translate along the line direction. 
The result is
a net motion in the climb direction perpendicular to the glide plane
(see Supplemental Material \cite{footnote2} for animations).
If no jogs are present, climb proceeds by a more
uniform simultaneous translation of larger line segments or
of the entire line, though the energy barrier for this type of climb is larger
than that of diffusive jog translation.

As shown previously for 2D triangular crystals in PFC \cite{pfcdisloc06,mpfc},
we find in 3D that
uniform climb velocities of undissociated dislocations in the limit of low
dislocation density ($\rho_d \lesssim 10^{13}m^{-2}$) 
follow a power law as a function of
applied stress, $v \sim \sigma^m$, with $m \sim 1$ for both dynamic equations,
Eqs.\ (\ref{eq:pfcdyn1}) and (\ref{pfcinertia}).
Apparent exponents $m$ as large as 4 can appear at 
higher dislocation densities.
Qualitatively similar behaviors as a function of $\rho_d$ 
have been observed in kinetic Monte
Carlo simulations \cite{kabirclimbKMC10}.

We also note that screw dislocation cross-slip readily occurs in 3D
PFC simulations, and we find that
the cross-slip barrier for dissociated FCC screw 
dislocations increases strongly with dissociation width and 
therefore with inverse stacking fault energy
(see Supplemental Material \cite{footnote2} for animations).
Such results, though far from a complete survey,
demonstrate that relatively simple 
PFC models are capable of capturing
many of the central atomistic features of plasticity in FCC and BCC materials
over diffusive time scales.

\section{Frank-Read-Type Glide Sources}
\label{sec:frsources}

The Frank-Read dislocation multiplication mechanism is 
an important element of crystal plasticity that has been widely observed
and studied both experimentally and via computer simulations
\cite{hirthlothe,kubinDDbook13,DDfrankreadalloys13,frankreadMDfcc03}.
In the most commonly discussed scenario,
a dislocation line pinned at two points within its glide plane bows out 
under stress until it meets itself on the opposite side of the pinning 
points. The contacting segments annihilate,
resulting in a single loop and the original pinned segment, which
then repeats the process if sufficient strain energy is still available.
The many possible manifestations of this basic mechanism,
including pinning of existing dislocation lines and emission of new
line segments from grain boundaries (which act as pinning centers),
contribute centrally to the large increases in dislocation density
that occur during plastic deformation.

Though Frank-Read mechanisms occur via conservative dislocation motion
and are therefore readily studied with
conventional MD 
\cite{frankreadMDfcc03}
and/or mesoscale continuum methods
\cite{kubinDDbook13,DDfrankreadalloys13},
we consider them here because of their general importance in plastic
deformation processes, to demonstrate that the basic physics of such
sources is well-captured by PFC models.
Non-conservative dislocation creation methods
are separately considered in the following section.
Unless noted otherwise,
all simulations presented in this section describe FCC materials and employ
Eqs.\ (\ref{pfcfree}), (\ref{pfcmod}), and (\ref{pfcinertia}).

\subsection{Prismatic sources}
\label{subsec:prismaticsources}

\begin{figure}[btp]
 \centering{
   \includegraphics*[width=0.48\textwidth,trim=70 456 70 87]{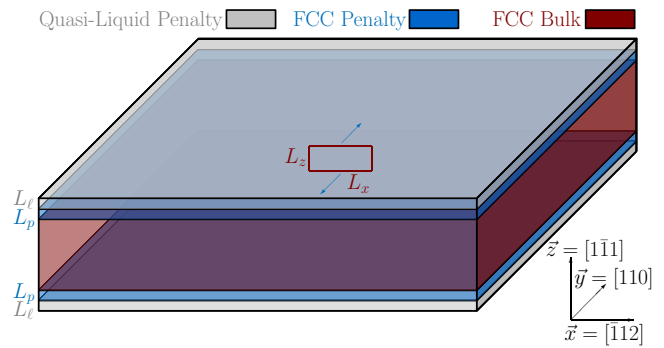}
 }
 \vspace{-.25cm}
\caption[]
{\label{FRschematic}
(Color online)
Schematic of the simulation setup used for Frank-Read source operation.
}
\end{figure}

\begin{figure}[btp]
 \centering{
   \includegraphics*[width=0.48\textwidth,trim=0 0 0 0]{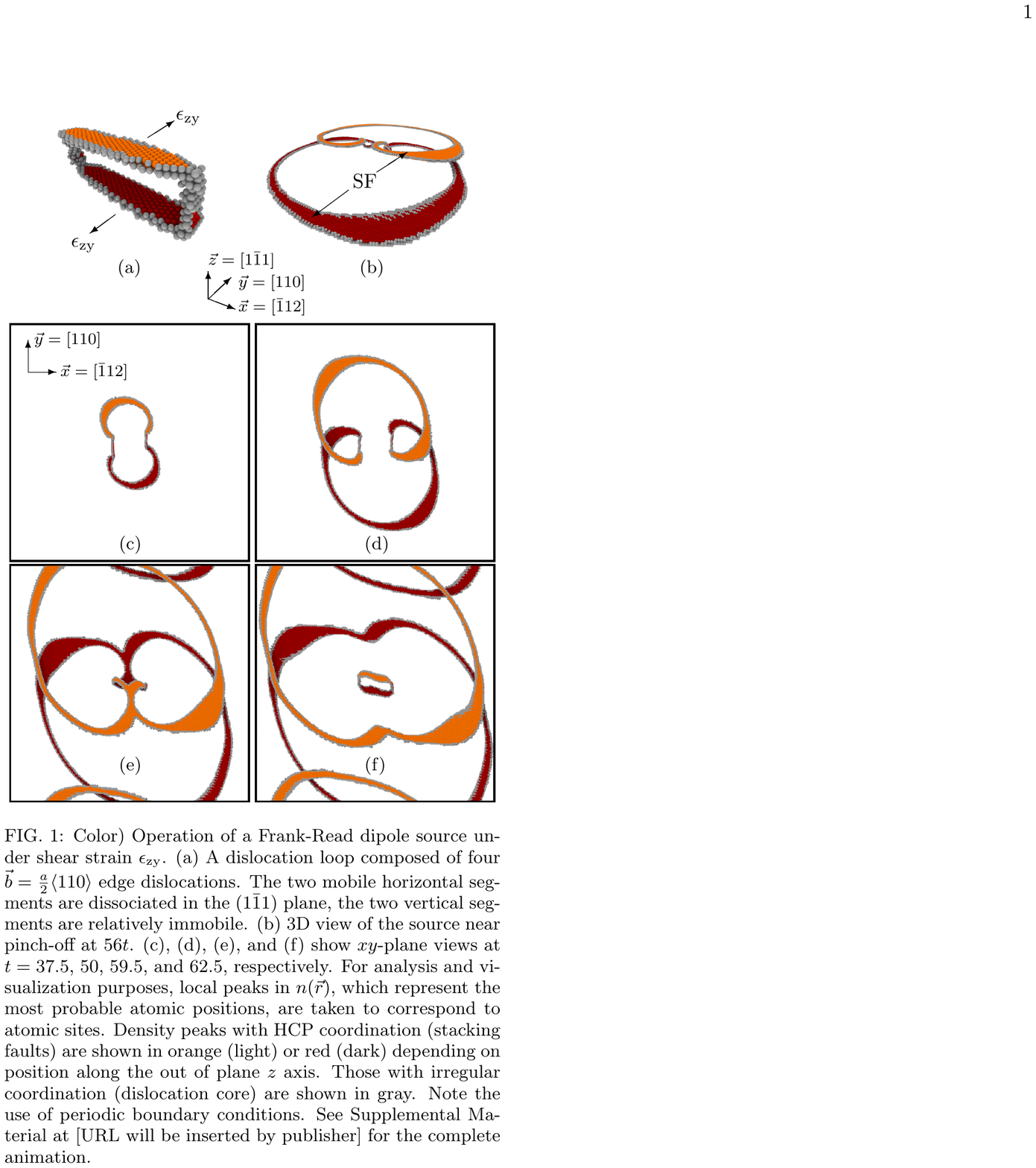}
 }
\caption[]
{\label{FRdipole}
(Color online)
Operation 
of a Frank-Read dipole source under shear strain $\epsilon_{\rm zy}$.
(a) A dislocation loop composed of four $a/2\langle110\rangle$ 
edge dislocations.
The two mobile horizontal segments are dissociated in the $(1\bar{1}1)$ plane, 
the two vertical segments are relatively immobile.
(b) 3D view of the source near pinch-off at $560t$.
(c), (d), (e), and (f) show $xy$-plane views at
$t=375$, $500$, $595$, and $625$, respectively.
For analysis and visualization purposes, 
local peaks in $n(\vec{r})$,
which represent the most probable atomic positions,
are taken to correspond to atomic sites.
Density peaks with HCP coordination (stacking faults) 
are shown in orange (light) or red (dark) depending
on position along the out of plane $z$ axis.
Those with irregular
coordination (dislocation core) are shown in gray.
See Supplemental Material \cite{footnote2}
for the associated animation.
}
\end{figure}

We first examine a Frank-Read dipole source consisting of a single
rectangular, prismatic edge dislocation loop in which the 
$a/2\langle110\rangle$ lines along 
the $[1\bar{1}1]$ direction are relatively immobile (sessile), while the
dissociated lines along the $[\bar{1}12]$ direction are mobile (glissile)
(see Figs.\ \ref{FRschematic} and \ref{FRdipole}a).
The $[1\bar{1}1]$ lines act as pinning points, such that
under applied shear strain $\epsilon_{\rm zy}$ the glissile lines
bow out via glide in opposite directions.
This setup was used in the MD simulations of 
Ref.\ \cite{frankreadMDfcc03}.
Such a loop could in principle be formed, for example, by vacancy agglomeration
following plastic deformation or irradiation, but it more generally
provides a convenient source configuration that is entirely analogous
to that of a longer jogged or locally pinned line segment.

To allow application of simple shear strain $\epsilon_{\rm zy}$,
the standard penalty function approach was used \cite{mpfc}.
The penalty function
is written as an additional free energy 
term of the form
$M(\vec{r})(n(\vec{r})-n_p(\vec{r}))^2$, where
$M(\vec{r})$ controls 
the strength of the penalty field
and $n_p(\vec{r})$ is the configuration of the penalty field.
We specify
\begin{eqnarray}
    M(\vec{r}), n_p(\vec{r}) = 
\begin{cases}
    M_{\ell}, n_0                      & \text{if } z \in L_{\ell}\\
    M_P, n^{\rm EQ}_{\rm FCC}(\vec{r}) & \text{if } z \in L_P\\
    0, 0                               & \text{otherwise}
\end{cases}
\end{eqnarray}
where $M_{\ell}$, $M_P$, $L_{\ell}$, and $L_P$ are constants
(see Fig.\ \ref{FRschematic})
and $n^{\rm EQ}_{\rm FCC}(\vec{r})$ is the commensurate equilibrium 
FCC $n(\vec{r})$.
The resulting system is a thin infinite slab of FCC bounded in the $z$-direction
by a homogeneous quasi-liquid phase of width $2L_{\ell}$.
The quasi-liquid layer simply circumvents any unphysical strains that 
would otherwise
be caused by the large shear disregistry at the periodic $z$ boundary.
Simple shear strain $\epsilon_{\rm zy}$ can then be applied to the
crystalline slab
by translating the upper pinned region $L_p$ along $+\vec{y}$ and
the lower pinned region $L_p$ along $-\vec{y}$ at some constant 
velocity.
When Eq.\ (\ref{pfcinertia}) is employed to allow rapid elastic 
relaxations,
a nearly uniform shear profile is produced across the sample. 

Various initial loop sizes and stacking fault energies were considered, with
two representative results shown in Figs.\ \ref{FRdipole} 
and \ref{FRquad}
(see Supplemental Material \cite{footnote2} for the associated animations).
In both cases, model parameters
$n_0=-0.48$, $r=-0.63$, $B^x=1$, $\alpha_0=1/2$, $k_0=6.2653$, 
$H_0=0.025$, $\beta=0.01$, and $\alpha=1$ were used.
Additional simulation details are given here \cite{footnote6}.
To estimate simulation time scales, we match the numerically measured 
FCC vacancy diffusion constant
$D_v \simeq 1.0 a^2/t$ to that of 
Cu at $1063^{\circ}$C 
($D_v \simeq 10^{-13}m^2/s, a \simeq 0.36nm$) \cite{pfcdisloc06}.
The 
time unit $t$ is then found to correspond to $\sim 1.3\mu s$,
and the shear rate $0.000235/t$ converts to $\sim 180/s$, which is roughly
five orders of magnitude lower than that of a typical and comparable 
MD simulation
(applying $1.8\%$ strain in $1ns$ produces a rate of 
$1.8\times10^7/s$ or see, e.g.\, Ref.\ \cite{rodneySFT06}).
Other FCC shear rates used in this study range from $\sim 10/s - 800/s$.

\begin{figure}[btp]
 \centering{
   \includegraphics*[width=0.48\textwidth,trim=0 0 0 0]{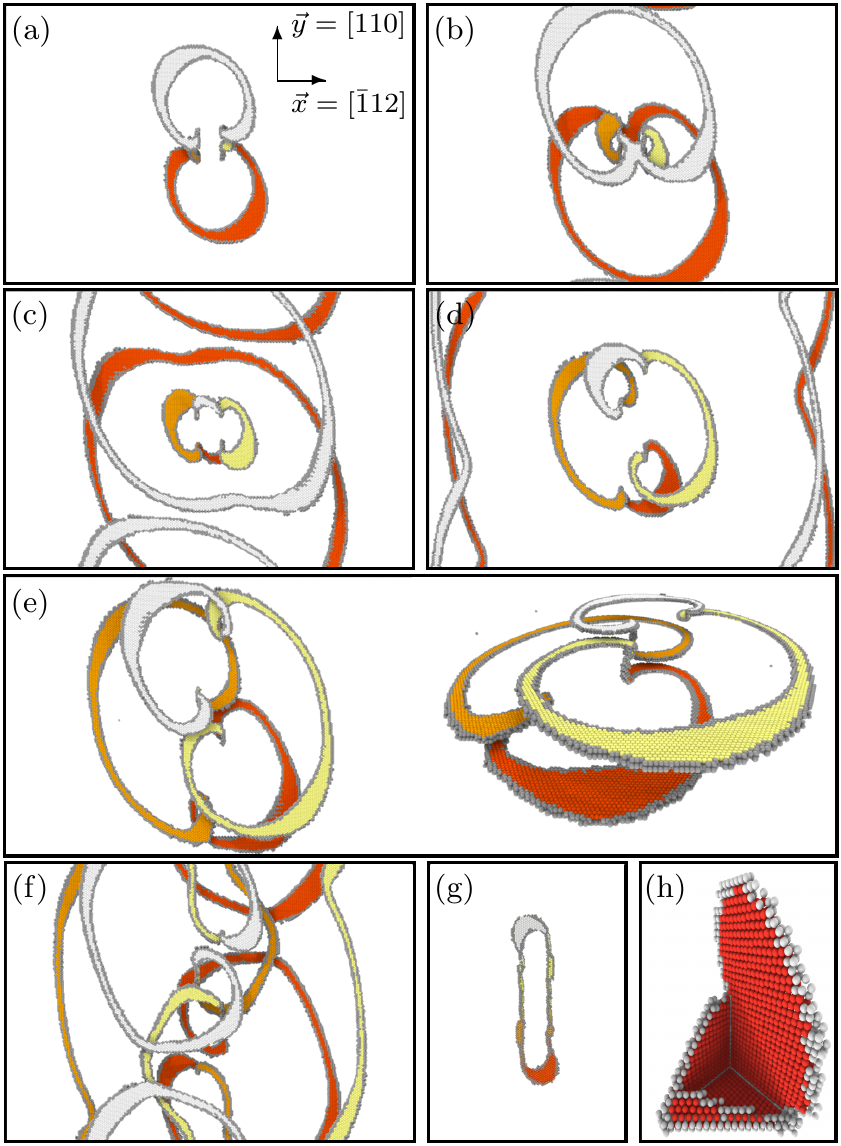}
 }
\caption[]
{\label{FRquad}
(Color online)
Operation 
of a Frank-Read dipole-to-quadrupole source.
The initial loop is shorter in the $x$-direction and longer in the
$z$-direction than that of Fig.\ \ref{FRdipole}.
(a), (b), (c), (d), (e), (f), and (g) show $xy$-plane views at
$t=450$, $520$, $580$, $680$, $750$, $840$ and $1030$, respectively.
A 3D view at $t=750$ is also shown in (e).
The source emits two initial loops near $520t$, then reorients
and converts into a quadrupole source before being restored to a
closed loop form near $1030t$.
Density peaks with HCP coordination (stacking faults) 
are shown in white, yellow (light gray), orange (gray), or red-orange
(dark gray), depending
on position along the out of plane $z$ axis.
Those with irregular
coordination (dislocation cores) are shown in gray.
(h) An unzipped stacking fault tetrahedron acting as a Frank-Read-type
source.
Density peaks with HCP coordination (stacking faults) 
are shown in red (dark gray), those with irregular
coordination (dislocation cores) are shown in gray.
See Supplemental Material \cite{footnote2} for the associated animations.
}
\end{figure}

The operation of the dipole source shown in Fig.\ \ref{FRdipole} closely
follows the basic Frank-Read mechanism, with slight asymmetries in loop
shape caused by image stresses in the $z$-direction and interactions
between opposing loops. 
Also, the $[1\bar{1}1]$ line segments, though sessile,
do not respond purely rigidly to local stresses, they are not 
perfectly pinned to their initial lattice locations. As the 
shear strain $\epsilon_{zy}$ increases and the Frank-Read loops begin
to bow out, the growing forces exerted on the $[1\bar{1}1]$
segments stretch these lines such their angle from vertical $\theta$ 
becomes larger than that of the applied shear $\theta_A=\tan{\epsilon_{zy}}$.
These small strain relief
mechanisms also contribute to the asymmetric shape of the growing loops,
and can lead to
more complex
effects when the prismatic loop 
dimensions
and strain rate are varied.

One such effect is shown in Fig.\ \ref{FRquad}.
Here the 
initial $[\bar{1}12]$ line segments are shorter and the
initial $[1\bar{1}1]$ segments are longer.
When the $[1\bar{1}1]$ segments stretch
under the influence of the loop bow-out stresses, they 
begin to approach alignment with the nearest of the four $\{111\}$ planes.
They are then able to lower their energy by taking a
periodically jogged configuration with dissociated 
glissile
segments in
adjacent $\{111\}$ planes, each connected by a single jog where the
lines are locally constricted.
After the maximum bow-out stress has been overcome and the growing loops become 
nearly circular (Fig.\ \ref{FRquad}a), the now jogged pinning
lines begin to hinge back toward vertical.
Rather than returning to their original undissociated, unjogged configuration,
the most highly strained segments near the loop ends 
cross-slip at their constricted jog sites onto the $(1\bar{1}1)$ glide
plane normal to the $z$-direction. These segments cannot easily cross-slip
back to the original configuration, they
instead begin to bow out on the $(1\bar{1}1)$ plane with the same jog sites
now serving as pinning points for the new sources.

Depending on $L_z$ and $\dot{\epsilon}_{zy}$,
the number of new intermediate loops 
nucleated can vary. In the case of Fig.\ \ref{FRquad}, two additional
loops are formed and the dipole is converted into a quadrupole.
The stresses exerted by the new loops cause the $[1\bar{1}1]$ lines
to reorient such that all segment pairs align with the shear direction
and the bow-out direction becomes perpendicular to the shear direction
(Fig.\ \ref{FRquad}d,e). Once all four loops are pinched-off, the
initial loop configuration is restored (Fig.\ \ref{FRquad}g).

If the maximum or activation stress of the initial source is low, then
the $[1\bar{1}1]$ segments may never reach the angle 
required to attain a 
stepped configuration
or may attain a stepped configuration with only very short glissile segments.
In such a case, no additional sources are formed.
As the maximum or activation stress of the initial source increases,
the length of the resulting glissile segments in the stepped configuration
also increases. If the length of any of these segments becomes large enough
to activate its operation as a new source for a given stress, then it will
bow out and begin forming a new loop. 
This type of transformation is therefore most likely to occur when such
sources have a large activation stress, which for Frank-Read 
\cite{hirthlothe} 
sources
$\sim 1/L_x$, corresponding to small $L_x$. 
Greater potential segmentation lengths (large $L_z$) also favor this behavior.
These expectations are in agreement with our results.

Such a transformation from dipole to multi-dipole also
requires that the pinning points have some small,
non-zero mobility, i.e., the pinning is not absolute.
This condition is perhaps more applicable to
soft crystalline materials such as colloidal crystals than to
metals, for example, but similar imperfect pinning behaviors 
have been reported in MD and atomistic quasicontinuum studies
of metallic crystals
\cite{MDlcstrength09,MDlcspirals12,junctionstrength99}.
Such transformations may therefore be observable in MD simulations.
Analogous situations should also become more
probable, for example, following a rapid quench or at high temperatures 
where vacancy concentrations are large and climb is active.

The general behaviors discussed in these two examples
exhibit some dependence on stacking fault energy $\gamma_{\rm ISF}$.
When $\gamma_{\rm ISF}$ is small, 
the strain required to operate the source tends to be lowest, and
the pinning segments remain relatively immobile throughout operation.
When $\gamma_{\rm ISF}$ is large, 
the strain required to operate the source increases, and
the forces exerted on the pinning segments by the bowing loops increase
such that the pinning segments may be dragged through the crystal,
effectively destroying the source.

It was also observed that stacking fault tetrahedra, created under low
$\gamma_{\rm ISF}$ conditions, can act somewhat similarly as multi-polar 
Frank-Read-type sources at high strains
\cite{footnote2}.
The $a/6\langle011\rangle$ Lomer-Cottrell (LC) 
stair-rod junctions that make
up the SFT edges are sessile, but under sufficient stress they may 
unzip and emit
$a/6\langle112\rangle$ Shockley partials that are pinned to 
neighboring
vertices of the tetrahedron. The emitted segments bow out and can eventually
form new, freely growing loops (see Fig.\ \ref{FRquad}h).
The observed critical strain for unzipping and activation of the leading
Shockley partial bow-out ($\sim \mu/25$) was found to
decrease as either $\gamma_{\rm ISF}$ or the SFT size is increased.

\subsection{Grain boundary sources}
\label{subsec:gbsources}

Dislocations can also be emitted from or absorbed into grain boundaries 
during plastic deformation. The initiation of this type of nucleation 
process is more structurally complex than those already discussed
due to the disordered nature of high angle grain boundaries, but 
the basic elements of the Frank-Read mechanism are still present.
If stress builds up near or within a grain boundary, it may be relieved
by grain boundary sliding or migration, for example, but in many cases
strain is most readily relieved by spontaneously nucleating new dislocation
lines which are then translated into the grain interior.
These lines may begin as small half-loops pinned to the boundary
at either end, which then bow out much like a Frank-Read source.
Rather than sweeping around the pinning points and pinching-off a complete
loop (the grain boundary generally prevents this)
the pinning points are more likely to migrate along the grain boundary 
as the half-loop grows.

Examples of such grain boundary nucleation processes in FCC
nanopolycrystals are shown in Fig.\ \ref{FR_GB}a-d
(see Supplemental Material \cite{footnote2} for the associated animation).
This system has the same model parameters as those of Figs.\
\ref{FRdipole} and \ref{FRquad}, and the grain structures were formed
using Voronoi tessellation.
Tensile strain $\epsilon_{\rm zz}$ was then applied at a constant rate
under constant volume conditions, such that 
$\epsilon_{\rm xx}=\epsilon_{\rm yy}=(\epsilon_{\rm zz}+1)^{-1/2}-1$.
Shear strain $\epsilon_{\rm zy}$ was found to produce similar results.
As a point of reference, the simulation shown,
which contains $\sim 1.15 \times 10^6$ atoms or density peaks, 
required 57 hours of wall-clock time to execute $2.4 \times 10^4$ time steps 
(or $\sim 3 ms$ estimated duration for Cu at $1063^{\circ}$C)
using 48 CPU cores.
A direct 
comparison with LAMMPS benchmark data for Cu with an EAM
potential \cite{lammpsbench} indicates
that the same computation of 
$\sim 1.15 \times 10^6$ atoms for
$3 ms$ on 48 CPU cores would require $\sim 1.25\times10^8$ hours of 
wall-clock time,
more than six orders of magnitude greater than the PFC time.

\begin{figure}[btp]
 \centering{
   \includegraphics*[width=0.48\textwidth,trim=0 0 0 0]{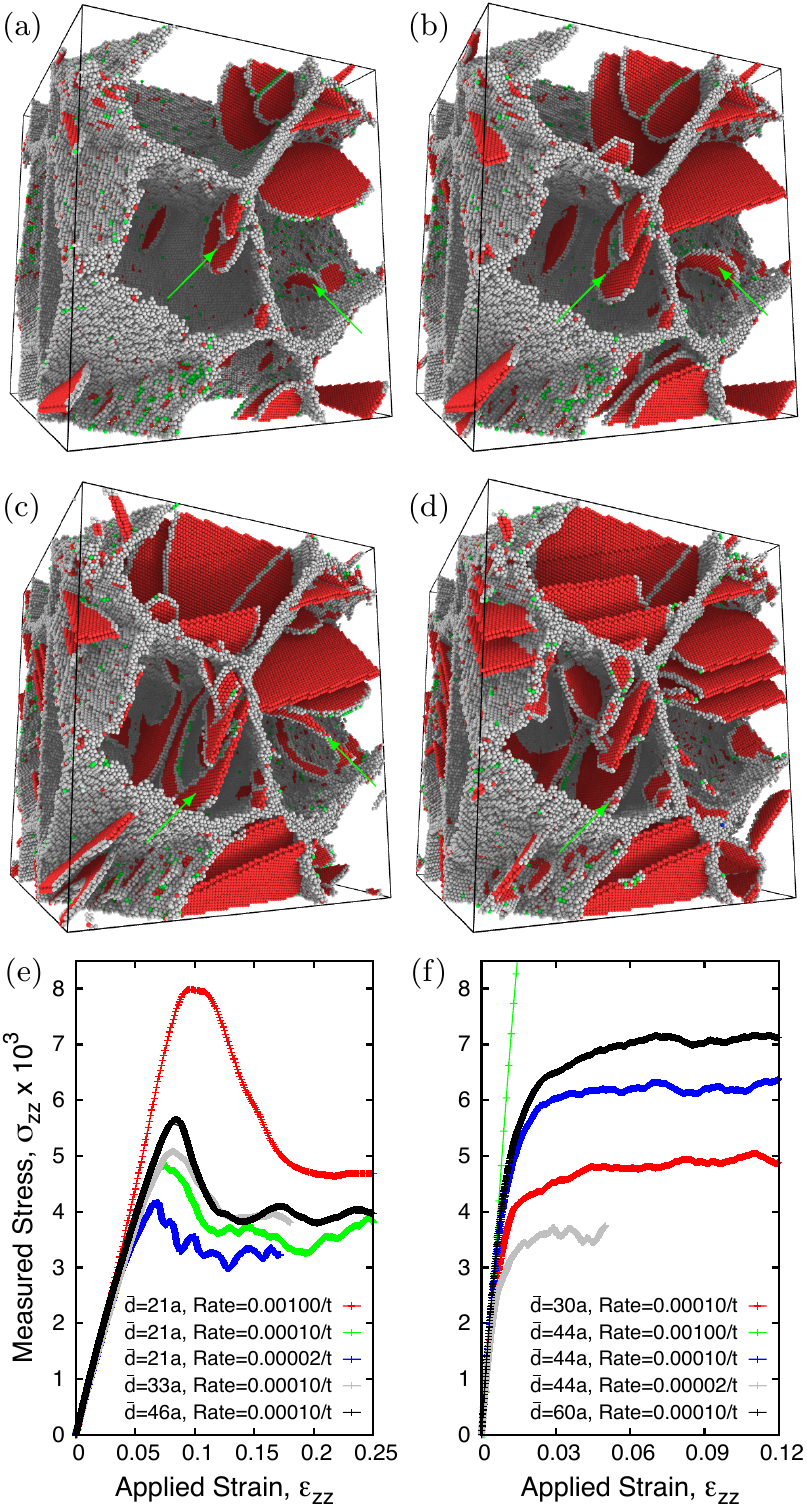}
 }
\caption[]
{\label{FR_GB}
(Color online)
Emission 
of dissociated $a/2\langle110\rangle$ 
dislocations
and $a/6\langle121\rangle$ partials from grain boundaries.
(a)-(d) Nanopolycrystalline FCC sample with average grain size
$\bar{d}\simeq 46a$ under tensile strain 
$\epsilon_{\rm zz}=0.082$, $0.086$, $0.092$, and $0.1$, respectively.
Both full dislocations (leading and trailing partials with stacking faults,
two of which are tracked by green (light gray) arrows) and leading partials
with stacking faults are emitted from / absorbed into grain boundaries, 
beginning near the yield point.
Density peaks with HCP (stacking faults), irregular (dislocation cores), 
and FCC coordination
are shown in red (dark gray), gray, and green (light gray), respectively.
(e) Stress-strain curves for FCC polycrystals with
various average grain sizes $\bar{d}$ and applied strain rates
(constant strain rate tensile deformation).
Model parameters are the same as those of Figs.\ \ref{FRdipole} and
\ref{FRquad}.
(f) 
As (e), but for BCC polycrystals 
with model parameters identical
to those reported in Section \ref{sec:bhsources}, except
$\alpha_1=2$, and Eq.\ \ref{pfcinertia} was employed with $\beta=0.01$,
$\alpha=1$.
See Supplemental Material \cite{footnote2} for the associated animations.
}
\end{figure}

Examples of the spontaneous nucleation of complete dissociated
$a/2\langle110\rangle$ dislocations are highlighted
with green arrows in Fig.\ \ref{FR_GB}.
These half-loops traverse the grain and are eventually absorbed into the 
opposite grain boundary as no fixed obstacles are present.
Numerous examples of leading partial nucleation and heavy faulting
are also apparent.
Each grain has four available $\{111\}$ planes in which the partials
may glide, resulting in faulting in some or all of these planes within
a given grain. 
The complex interactions of the various partials and
stacking faults lead to varying intra-grain textures and structures
with results very similar to those produced by MD simulations
and consistent with experimental observations
\cite{dislocsolids14,dislocsolids15}.

The stress-strain curves shown in Fig.\ \ref{FR_GB}e also confirm this
qualitative agreement. 
Stresses $\sigma_{\rm ij}$ were periodically
quantified by measuring the rate of change in average free 
energy of an instantaneous
$n(\vec{r})$ configuration as the appropriate deformation is statically applied.
For example, $\sigma_{\rm zz}$ was measured by varying the grid spacing
in the $z$-direction and quantifying 
$\sigma_{\rm zz} = \delta \bar{F}/\delta \epsilon_{\rm zz}$
in the small strain limit every ten time steps.
Upper and lower yield points are observed for FCC systems
with both yield stresses decreasing with decreasing rate,
as observed in MD studies \cite{dislocsolids14}.
The upper yield point is associated with initiation of plastic flow /
dislocation nucleation within the initial clean grains, while the
lower yield point is associated with steady-state 
dislocation and grain boundary driven plasticity within
the resulting dislocated systems.
Larger grain sizes produce higher yield points, indicating that
these systems are in the reverse Hall-Petch regime \cite{meyers2006review}.
This dominance of dislocation nucleation, glide, and annihilation,
and a lack of visible pile-ups suggests that 
this reverse Hall-Petch
behavior is associated with an absence of pile-up-induced hardening at
these small grain sizes.
Other mechanisms such as grain boundary sliding and dislocation source 
starvation may contribute to this behavior as
well, but we have not quantified the contributions of such effects.
We only note that they are visually less evident than the described
dislocation activity.

Analogous simulations of BCC polycrystals produced the results shown in 
Fig.\ \ref{FR_GB}f
(see Supplemental Material \cite{footnote2} for the associated animation).
A simpler yielding behavior is observed in this case,
as the BCC nanopolycrystals within this parameter range deform plastically
via grain boundary migration. Very little dislocation nucleation occurs,
leaving essentially a network of sliding and creeping grain boundaries.
Grain boundary mechanisms 
therefore appear to be entirely responsible for
the inverse Hall-Petch behavior in these systems.
Such behavior is consistent with
intermediate/high temperature experiments on nanopolycrystalline BCC
metals in which grain boundary mechanisms are found to dominate plasticity
\cite{dislocsolids15}.
The greater dependence of BCC stress-strain behavior on strain rate 
also indicates a larger diffusive creep component than in FCC.
This is in agreement with general observations of
higher creep rates in BCC materials, an effect ascribed to the
higher self diffusivity of non-close-packed structures \cite{defmapsbook82}.
These simulations of FCC and BCC polycrystals demonstrate that the 
qualitative features of conservative
dislocation emission from grain boundaries and of overall stress-strain 
response in elastic and plastic regimes can be well-captured by PFC models.

\section{Bardeen-Herring-Type Climb Sources}
\label{sec:bhsources}

Climb-mediated or Bardeen-Herring-type sources can become active at high
temperatures or following rapid quenches when the excess vacancy
concentration is large.
The basic principles are in many ways analogous to those of Frank-Read
sources, except that the dislocation motion is mediated by vacancy
diffusion rather than by glide.
Nonetheless such sources
are not as well understood as Frank-Read sources, at least partly because 
they are not easily
modeled at the atomistic level using conventional approaches, and
because mesoscale models must account for the often complex nature
of vacancy diffusion around dislocation cores and among other
heterogeneous strain fields.
We show here that PFC simulations permit the study of such sources
with atomistic resolution, and
reveal a range of complex nucleation
behaviors caused by non-trivial interactions between interface structure,
strain orientation, and dislocation energetics.

\subsection{Critical strain for loop nucleation}
\label{subsec:bhstrain}

The specific phenomenon considered in this section is nucleation
of dislocation loops from spherical objects such as precipitates,
inclusions, or voids in BCC crystals under uniaxial tension or compression
$\epsilon_{ii}$.
The case of loop nucleation and coherency loss at
precipitates has been studied in early
theoretical and experimental work 
\cite{precipnucl1968,weatherly1968,ashbystress1969,brownwoolhouse1970},
as precipitate coherency can have a significant impact on the
mechanical properties of metal alloys.
Concentric dislocation loops centered on precipitates or impurities
have also been observed in various metals \cite{concentric85,
concentric62,embury1963,concentric66,concentric89}.
These may be formed by Bardeen-Herring-type mechanisms similar to
those described here.
To our knowledge this problem has not been examined via
numerical simulations nor at the atomistic level due to the
long, diffusive time scales involved.

As strain is applied to a system containing a spherical precipitate, 
the free energy
eventually becomes higher than that of the same system with a
dislocation loop that is able to grow and relieve strain energy.
An energy barrier for the nucleation of such a loop will generally exist
such that its appearance in a dynamic simulation may be delayed to higher
strains. Nonetheless, a loop eventually appears at the sphere-matrix
interface, and when the applied strain is purely uniaxial, the nucleation
and growth of the loop is largely mediated by climb. Other strain types
can lead to different, relatively well-characterized conservative loop 
formation processes such as prismatic punching
\cite{balluffi05kinetics,ruddvoidpunching09,Calhounloops99}.

In a linear elastic isotropic continuum,
the critical strain $\epsilon^*_{ii}$ at which loop nucleation becomes
favorable is approximately
\be
\epsilon^*_{ii} = 
\frac{b}{8\pi(1-\nu)(\frac{B_S}{B_M}-1)R_0}
\left[ \ln{\frac{8R_0}{b}}+\frac{2\nu-1}{4(1-\nu)} \right]
\label{60stheory}
\ee
where $b$ is the magnitude of the dislocation Burgers vector,
$\nu$ is the Poisson's ratio of the matrix,
$B_S$ and $B_M$ are the bulk moduli of the sphere and matrix phases, 
respectively, and $R_0$ is the sphere radius 
\cite{precipnucl1968,brownwoolhouse1970}.
We note that anisotropy may play a role in the BCC system studied here.

Homogeneous spherical inclusions were introduced into the present simulations 
by adding a uniform penalty function over some predefined spherical volume
in the center of a simulation cell with initially perfect BCC crystal 
structure (see Fig.\ \ref{BHexample}). 
The resulting spherical body has a larger elastic modulus than that
of the bulk crystal, but since $n(\vec{r})$ within the sphere is uniform 
rather than periodic, 
the issue of coherent vs.\ incoherent interface structure is not relevant.
After the system was equilibrated, uniaxial strain $\epsilon_{ii}$
was applied at a
constant rate by uniformly increasing or decreasing the numerical grid
spacing along one axis of the periodic simulation cell by a small
amount at every time step.

Equations (\ref{pfcfree}), (\ref{xpfckernel}), and (\ref{eq:pfcdyn1}),
were employed in all simulations discussed in this section.
Parameter values used were
$w=1.4$, $u=1$, $n_0=0$, $\alpha_1=1$, $\sigma=0.1$,
$\rho_1=1$, and $\beta_1=8$.
Other simulation details are given here \cite{footnote10}.
Following the procedure used for FCC simulations,
if we match the numerically measured BCC vacancy diffusion constant
$D_v \simeq 1.5 a^2/t$ to that of vanadium at $1842^{\circ}$C
($D_v \simeq 1.36\times10^{-13}m^2/s, a \simeq 0.302nm$),
then the time unit $t$ is found to correspond to $\sim 1\mu s$.
The range of shear rates used in BCC simulations then converts to
$\sim 20/s-2500/s$, values again roughly
four to six orders of magnitude lower than those of typical and comparable 
MD simulations.

\begin{figure}[btp]
 \centering{
   \includegraphics*[width=0.48\textwidth,trim=0 0 0 0]{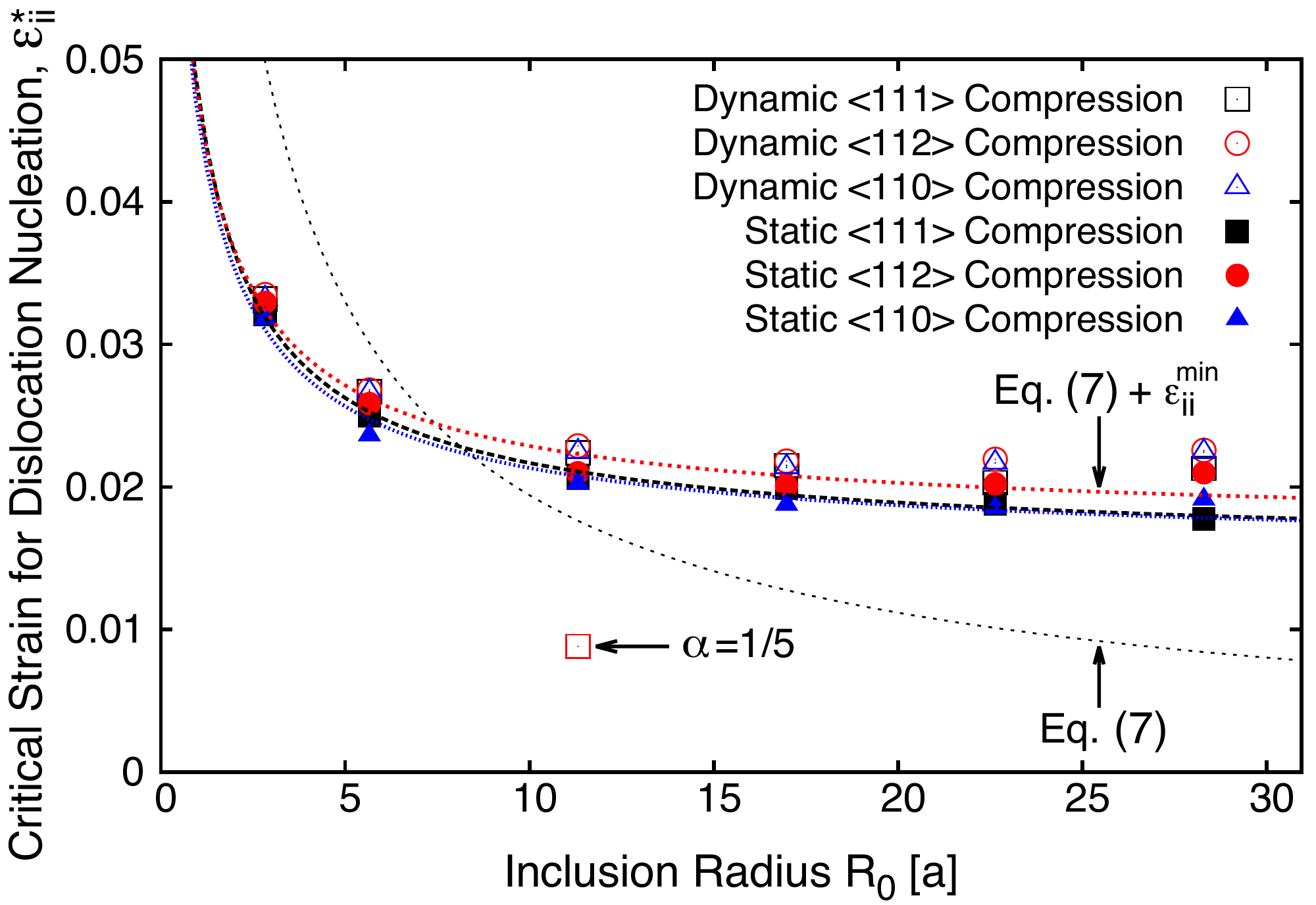}
 }
\caption[]
{\label{BHplot}
(Color online)
Critical strain $\epsilon^*_{ii}$ for Bardeen-Herring
climb source activation vs sphere radius $R_0$.
The points represent simulation data at different effective strain rates,
and the lines are predictions of the theory of Brown {\it et al}.\
\cite{precipnucl1968}, without and with a finite minimum critical strain.
The fits employ fixed parameters
$b = 1/\sqrt{2}$ and $\nu=1/3$,
and adjustable parameters
$\epsilon^{\rm min}_{ii}$ and $B_S/B_M$,
which are in all cases close to $0.016$ and $4$, respectively.
}
\end{figure}

Compiled results for the critical nucleation strain 
$\epsilon^*_{ii}$ at all sphere sizes and strain orientations
are shown in Fig.\ \ref{BHplot}.
The agreement between the low strain rate results and the static energy
minimization results indicates that any rate dependence is minimal
at the slower rate considered.
The general trend is a decrease in $\epsilon^*_{ii}$ as $R_0$ increases.
The form of the decrease is well described by Eq.\ (\ref{60stheory})
after an additional constant strain $\epsilon^{\rm min}_{ii}$ 
is added to the right hand side.
This constant is discussed further in the following paragraph.
The adjustable parameters in the fits shown are therefore
$\epsilon^{\rm min}_{ii}$ and $B_S$, as there are some ambiguities
in the effective value of the bulk modulus of the sphere as modeled.
Nonetheless, the fits are quite good for $B_S/B_M \simeq 4$,
which seems to be a reasonable estimate of the ratio produced by
our simulations.

It is not conclusive whether $\epsilon^*_{ii}$ will continue to slowly decrease
as $R_0$ becomes very large or whether it levels off to some minimum value.
Our expectation is that $\epsilon^*_{ii}$ will plateau in the PFC
simulations due to finite size effects as well as the existence of a threshold
Eckhaus-like strain for activation of the wavelength selection or climb
`instability' \cite{pfcbook2011}.
Both of these effects are driven by lattice periodicity.
An integer number of unit cells must fit into the simulation box,
such that the energy of a perfect crystal will not be reduced by
the addition or removal of a plane of atoms or density peaks until
$\epsilon_{ii} > 1/(2N_i)$, where $N_i$ is the number of unit cells
in the $i$ direction.
Furthermore, there will be an energy barrier for this removal process
associated with an Eckhaus-like instability strain, which quantifies the
strain at which this barrier goes to zero \cite{pfcbook2011}.
Thus, without thermal fluctuations we expect to observe
$\epsilon^*_{ii} > 1/(2N_i) \simeq 0.009$
for the system size used in this study. 
As the elastic
moduli become large, $\epsilon^*_{ii} \rightarrow 1/(2N_i)$
since the perfect crystal Eckhaus strain is roughly proportional to $\alpha_i$.
The single data point at $\alpha_1=1/5$, $R_0=15.8b$ is consistent
with this expectation.

\subsection{Loop geometry and evolution}
\label{subsec:bhgeo}

The critical strain for nucleation is therefore in general agreement
with continuum elastic predictions, but it is worthwhile to examine
the nucleation and growth process in greater detail.
A typical result from the dynamic simulations is shown in 
Fig.\ \ref{BHexample}
(see Supplemental Material \cite{footnote2} for the associated animation).
The sphere radius in this case is $R_0=11.3a$, and the strain is
compressive along the $x$ axis, $\epsilon_{\rm xx}$.
A loop first begins to form at the sphere-matrix interface with
a slightly serpentine shape due to the variations in local line energy
around the surface of the sphere.
Essentially, the nature in which the spherical surface intersects
the various crystallographic planes of the matrix creates
a quasi-2D energy landscape on the spherical surface
which the dislocation loop must navigate to minimize its total energy
with the constraint of fixed total Burgers vector.
Certain planes and line directions will be preferred over others.
This effect involves not only crystallographically-dependent dislocation
energies, but also atomic-level core structure effects as well as
elastic anisotropy,
which together are beyond the scope of continuum elastic theories.
The impact on initial loop shape tends to be small for small $R_0$ but 
increases considerably for larger $R_0$ values, as will be shown.

\begin{figure}[btp]
 \centering{
   \includegraphics*[width=0.48\textwidth,trim=0 0 0 0]{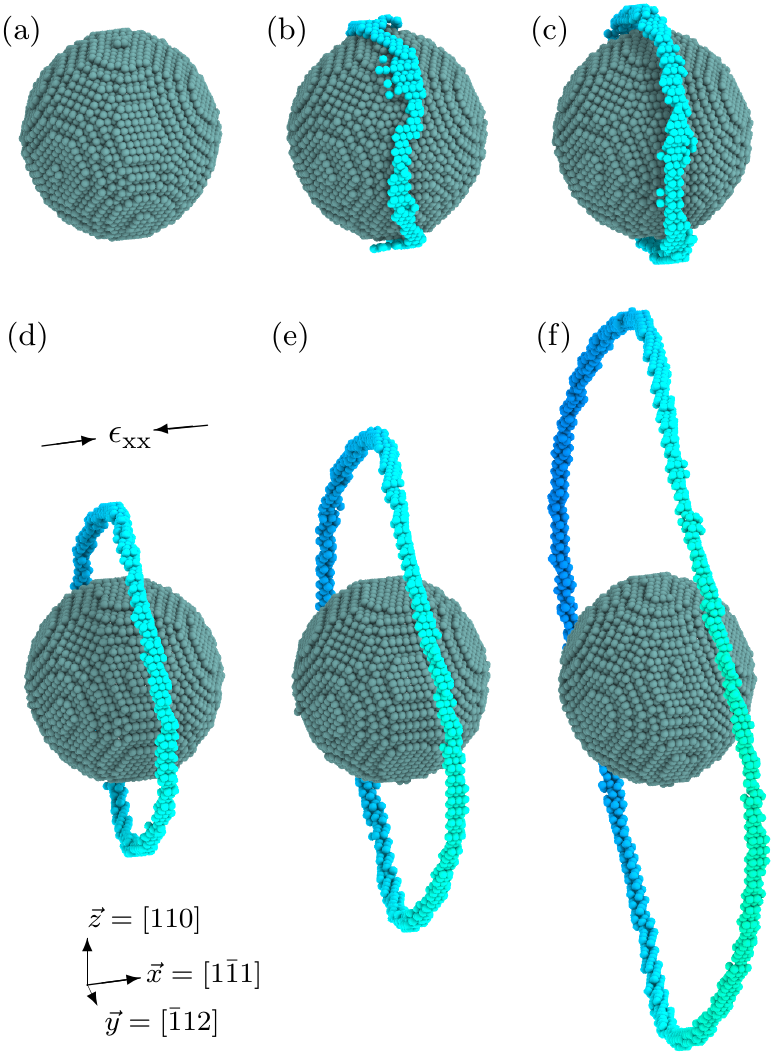}
 }
\caption[]
{\label{BHexample}
(Color online)
Operation 
of a Bardeen-Herring-type spherical climb source with $R_0=11.3a$
in a BCC crystal under uniaxial compression $\epsilon_{\rm xx}$.
Time evolution shown at (a) $t=110$, (b) $t=126$, (c) $t=135$, 
(d) $t=145$, (e) $t=160$, and (f) $t=180$.
Only density peaks with irregular coordination (interface and dislocation core 
sites) 
are displayed. Those at the sphere-matrix interface are shown in gray (dark),
those inside the dislocation core are shown in blue-green (light) depending on
position along the $y$-axis.
See Supplemental Material \cite{footnote2} for the associated animation.
}
\end{figure}

The initial loop nucleation and its subsequent growth both require
vacancy diffusion to or away from the 
surface
{\color{black}
of the sphere.
In the PFC approach, this process is mediated by local diffusive
modes of the density wave amplitudes.
After the loop detaches from the sphere, its 
shape continues to evolve
as the effective local energy landscape changes with loop radius.
Non-planar, non-circular shapes are common as the competition between
minimum static energy and lowest energy pathway to continued growth
can be delicate and non-trivial. This is somewhat analogous to
the cross-slip of dissociated screw 
dislocations, during which local
segments must constrict at large energy cost to permit cross-slip
into the next 
local energy minimum. Any instantaneous configuration may
not be the lowest energy static configuration for the given loop radius, 
but it should facilitate evolution toward an even lower energy state
with larger radius.
}

A more complex loop nucleation process is shown in Fig.\ \ref{BHexample2}
(see Supplemental Material \cite{footnote2} for the associated animation).
The only difference from Fig.\ \ref{BHexample} is the increased sphere
size of $R_0=28.3a$.
The result is a much more pronounced serpentine shape due to the
larger areas on the surface of the sphere that nearly coincide with
low energy lattice planes of the matrix.
Within the $(110)$ plane for example, the preferred line direction
for the nucleated $a/2[1\bar{1}1]$ dislocation is  
along the nearest $\langle001\rangle$ vector rather than the 
$[\bar{1}12]$ vector normal to the strain axis. 
The dislocation therefore forms with mixed edge-screw character 
along the $[001]$ direction within the uppermost $(110)$ plane for example, 
taking the loop locally out of alignment
with the strain-normal $(1\bar{1}1)$ plane. The line must then wind back
toward the strain-normal plane in the areas where it curves out of
the top and bottom $(110)$ planes.

\begin{figure}[btp]
 \centering{
   \includegraphics*[width=0.48\textwidth,trim=0 0 0 0]{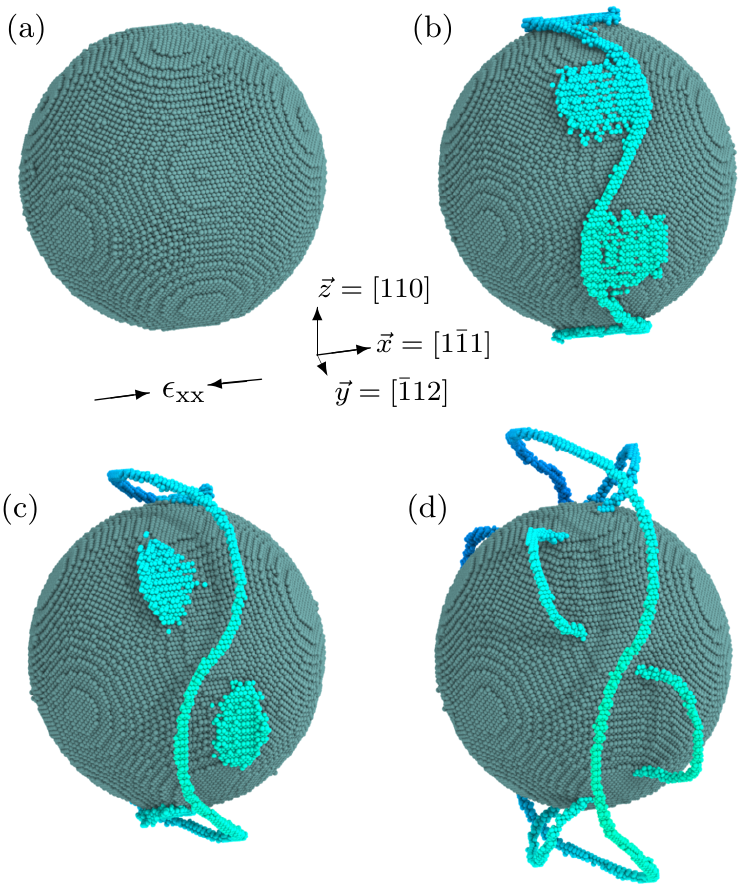}
 }
\caption[]
{\label{BHexample2}
(Color online)
Operation 
of a Bardeen-Herring-type spherical climb source with $R_0=28.3a$
in a BCC crystal under uniaxial compression $\epsilon_{\rm xx}$. 
Time evolution shown at (a) $t=110$, (b) $t=138$, (c) $t=150$, and (d) $t=162$.
Only density peaks with irregular coordination (interface and dislocation core 
sites) 
are displayed. Those at the sphere-matrix interface are shown in gray (dark),
those inside the dislocation core are shown in blue-green (light) depending on
position along the $y$-axis.
See Supplemental Material \cite{footnote2} for the associated animation.
}
\end{figure}

The same arguments hold as the line crosses through the other
intersecting $\{110\}$ planes.
The $a/2[1\bar{1}1](011)$ segments, for example,
prefer alignment with the $[\bar{1}00]$ direction. 
The $a/2[1\bar{1}1](\bar{1}12)$ segments on the other hand
prefer alignment with the $[110]$ direction (pure edge character).
Thus a serpentine winding pattern is produced as a result of the
dynamic competition between loop energy minimization, which tends to 
promote a certain degree of winding,
and strain relief maximization, which tends to suppress winding
in favor of maximizing the outward growth (climb) velocity.

Detachment and growth away from the sphere occurs first in regions
that do not align with any low energy dislocation slip planes.
Segments in low energy $\{110\}$ planes are observed to detach last,
as these have the lowest local energy and the highest barrier for out of
plane motion.
Terrace sites at the edges of faceted low energy planes in 
{\color{black}
particular
appear to have the maximum detachment barrier.
The extra half loops protruding from the sphere in Fig.\ \ref{BHexample2}
are a dynamic effect that disappears at low applied strain rates.
}

Other variations of climb-mediated loop nucleation processes
are shown in Fig.\ \ref{BHexample3}
(see Supplemental Material \cite{footnote2} for the associated animations).
Uniaxial strain along a $\langle110\rangle$ direction
produces two disjointed $a/2\langle111\rangle$ half-loops
that grow symmetrically at a $12.5^{\circ}$ angle to the strain-normal
$(110)$ plane (Fig.\ \ref{BHexample3}a).
An $a\langle100\rangle$ edge line segment appears at
the intersection of these half-loops, and can link to a second, concentric
inner pair of half loops that allows complete detachment from the sphere.
Uniaxial strain along a $\langle112\rangle$ direction
produces two separate $a/2\langle111\rangle$ edge
dislocation half-loops, as displayed in Fig.\ \ref{BHexample3}b.
These half-loops climb until their terminal ends meet and merge into a
single, nearly circular loop.
Dual loop nucleation as shown in Fig.\ \ref{BHexample3}c
is also possible for certain $R_0$, elastic moduli, and strain rates.
Finally, arrays of spheres simultaneously nucleate complex networks of
dislocation lines via mixed climb-glide processes.
Effects from dislocation-dislocation, dislocation-grain boundary,
and dislocation-inclusion interactions, for example, are naturally
incorporated into the evolution of such networks in PFC simulations.

\begin{figure}[btp]
 \centering{
   \includegraphics*[width=0.48\textwidth,trim=0 0 0 0]{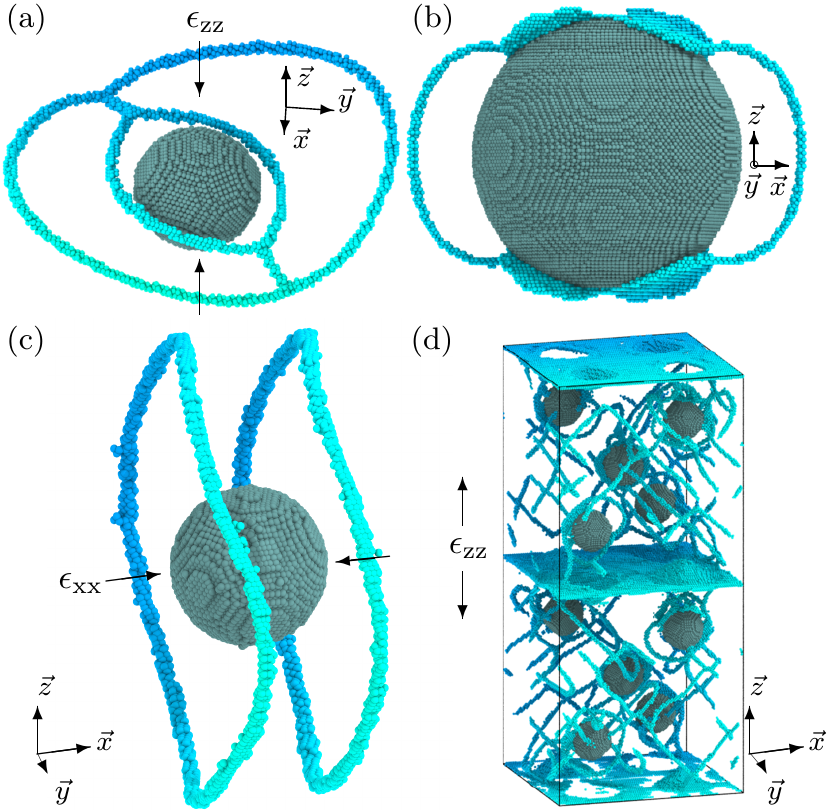}
 }
\caption[]
{\label{BHexample3}
(Color online)
Examples of various other Bardeen-Herring-type climb sources 
in BCC crystals.
(a) Linked, concentric, biplanar loop pairs under $\epsilon_{\rm zz}$
with $R_0=11.3a$.
(b) Offset nonconcentric loop pairs under $\epsilon_{\rm yy}$
with $R_0=28.3a$.
(c) Offset concentric loop pairs under $\epsilon_{\rm xx}$
with $R_0=9a$ and $\alpha=2$.
(d) Dislocation network / tangle formation from an array of 10 sources 
with $R_0=9a$ in a bicrystal under constant volume tension $\epsilon_{\rm zz}$.
Only density peaks with irregular coordination (interface and dislocation core 
sites) 
are shown. Those at the sphere-matrix interface are shown in gray (dark),
those inside the dislocation core are shown in blue-green (light) depending on
position along the out of plane axis.
See Supplemental Material \cite{footnote2} for the associated animations.
}
\end{figure}

\section{Dislocation-SFT Interactions}
\label{sec:dislocSFT}

Impediments to dislocation motion, including other 
dislocations,
planar faults, and 3D obstacles, play a central role in the mechanical
response and work-hardening 
properties of metals.
The dominant irradiation-induced defect in FCC materials is the SFT,
and dislocation-SFT interactions are therefore believed to largely control 
the mechanical response of 
FCC materials in nuclear applications
\cite{wirthSFT09}.
Such interactions have been widely studied in MD (and DDD) 
simulations
\cite{rodneySFT06,wirthSFT09,niewczasSFT09a}
providing for our purposes a potentially useful body of benchmark
results with which PFC 
simulations can be compared.
A few selected results 
are presented in this section.

\begin{figure}[btp]
 \centering{
   \includegraphics*[width=0.48\textwidth,trim=0 0 0 0]{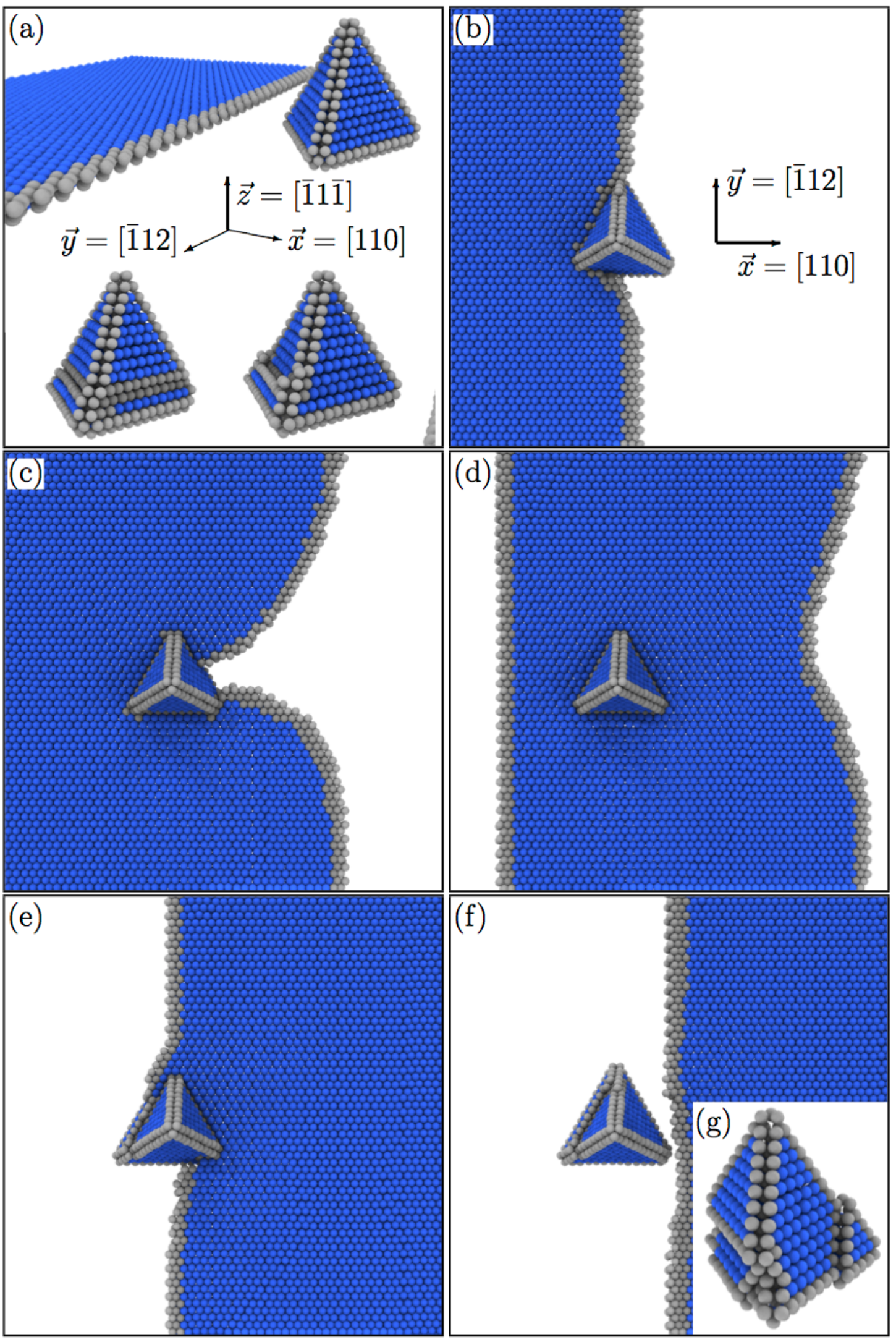}
 }
\caption[]
{\label{SFTedge}
(Color online)
Interaction between a SFT and a gliding dissociated edge dislocation in a
FCC crystal, 
($ED/Down, 4/13, 0.0001/t$).
Perspective views of the SFT and dislocation are shown in (a).
The upper image shows both at $t=10100$, 
while the lower left and right images
show the SFT at $t=15800$ and $t=20100$, respectively.
Only the leading partial has passed the SFT at $t=15800$, both partials have
passed by $t=20100$.
$xy$-plane views are shown in (b), (c), (d), (e), and (f) at
$t=12100$, $15100$, $16600$, $18600$ and $19600$, respectively.
(g) The damaged SFT following the ($SD/Edge, 4/13, 0.0001/t$) interaction.
Density peaks with HCP coordination (stacking faults) 
are shown in blue (dark gray), those with irregular
coordination (dislocation cores) are shown in gray (light gray).
See Supplemental Material \cite{footnote2} for the associated animations.
}
\end{figure}

SFTs were formed in the present simulations via the Silcox-Hirsch mechanism
\cite{silcoxhirschSFT59}.
A triangular Frank loop is initiated on a $\{111\}$ plane, after which it
spontaneously relaxes into the local energy minimum corresponding
to a perfect SFT with base prescribed by the initial Frank loop
(see Supplemental Material \cite{footnote2} for the associated animation).
Dissociated $a/2\langle110\rangle$
edge or screw dislocations were equilibrated some lateral
distance away from the SFT and some vertical distance relative to the 
SFT base. Shear strain $\epsilon_{zx}$ 
was then applied at a constant rate to cause the
dislocation to glide toward and through the SFT
(see Fig.\ \ref{SFTedge}).

Only two cases will be reported here, given in the notation of
Ref.\ \cite{rodneySFT06} as
($ED/Down, 4/13, 0.0001/t$) and
($SD/Edge, 4/13, 0.0001/t$).
In the first case, this notation indicates that an edge dislocation ($ED$) 
intersects a 
SFT with apex oriented in the $[\bar{1}1\bar{1}]$ direction 
($Down$), at 
the fourth $\{111\}$ plane from the base of the SFT
which is $13$ $\{111\}$ planes tall ($4/13$), and with shear rate $0.0001/t$. 
The second case is the same except that it considers a screw dislocation ($SD$) 
intersecting a SFT with one edge oriented along the $SD$ line direction
($Edge$).
Additional simulations details are given here \cite{footnote15}.
Results from the ($ED/Down, 4/13, 0.0001/t$) simulation are shown in 
Fig.\ \ref{SFTedge}.
The accompanying animation, along with that of
the SFT-screw dislocation interaction, can be found in the Supplemental
Material \cite{footnote2}.

The 
general sequence of events includes dislocation 
pinning at SFT Lomer-Cottrell stair rod junctions,
bow out of the gliding dislocation line between the image SFTs,
Orowan looping of the SFT by the leading partial, 
and damage of the SFT after the trailing
partial has passed through.
The pinning and bow-out effects are of course expected.
The Orowan loop created by the leading partial is also consistent with
MD simulations of individual Shockley partial-SFT interactions
\cite{niewczasSFT09a}. In the present simulations, 
the trailing partial eventually shears 
the SFT
as well, clearing the Orowan loop and leaving either one or two ledges on the
SFT faces. 
The ledge structures appear to be consistent with those observed in
MD \cite{rodneySFT06}.
The height of the SFT apex above is also reduced by one 
$\{111\}$ interplanar distance in the $ED$ case.

We have not yet observed other possible outcomes reported in MD simulations, 
such as partial SFT absorption and jog formation, but we have
considered only a very small subset of the conditions examined via MD.
We therefore argue that these results provide partial but strong
qualitative evidence that PFC simulations can correctly reproduce
complex defect phenomena of this type.

\section{Conclusions}
\label{sec:conclusions}
Basic dislocation properties in FCC and BCC crystals have been examined
in the context of phase field crystal models, and extended into simulations
of
conservative and nonconservative dislocation creation mechanisms 
and obstacle flow processes. 
Core structures of dissociated $a/2\langle110\rangle$ 
FCC dislocations and $a/2\langle111\rangle$ BCC dislocations
have now been reproduced in PFC with sufficient accuracy to capture many
aspects of plastic flow that derive from such structures.
These include the known anisotropy in BCC screw-edge glide mobility
as well as the effect of FCC dissociation width on cross-slip and 
climb barriers.
Classical Frank-Read-type sources have been simulated for the first
time with such models, and a new mechanism by which dislocation
lines or superjogs under strain can segment onto multiple glide planes,
converting local monopole or dipole sources into multipole sources,
has been identified.
Stacking fault tetrahedra under high strain have also been shown
to reconstruct 
{\color{black}
and emit dislocations via a Frank-Read-type mechanism.
Basic features of 3D polycrystal plasticity and dislocation emission from 
grain boundaries have also been examined and shown to be consistent
with MD simulation results.
}

Nonconservative dislocation creation mechanisms associated with
spherical precipitates, inclusions, or voids have been studied for the
first time using atomistic simulations.
Results for the critical strain to nucleate a loop from a spherical body
are in agreement with predictions of continuum elastic theory after 
accounting
for finite size effects
and moduli-dependent climb barriers present in our simulations.
A range of complex nucleation
behaviors caused by non-trivial interactions between interface structure,
strain orientation, and dislocation energetics have been revealed.
Observed loop geometries have been rationalized for a few select
cases, but the results in general highlight the 
sometimes unexpected complexity that
can emerge when atomistic effects associated with crystal structure,
dislocation cores, and climb dynamics are simultaneously considered.

The Silcox-Hirsch SFT formation mechanism has also been reproduced, 
as well as qualitative features of SFT-dislocation interactions observed in
MD and DDD simulations.
Such processes will require further study to gain a fuller
understanding of the similarities and differences between PFC and other
atomistic simulation methods. But there appears to be promise in the possibility
of simulating features of obstacle flow involving, for example,
climb bypass mechanisms that cannot be accessed with other conventional
methods.

In a wider sense, it is hoped that these results convey the potential
of the PFC approach as applied to solid-state materials phenomena in three
dimensions.
This type of description unifies conservative and nonconservative
plastic flow mechanisms with atomistic resolution, enabling
the study of complex high temperature diffusive evolution processes 
in the nanoscale size regime. Many such processes are
inaccessible to conventional atomistic approaches.
Applications to pure or multicomponent systems and phenomena such as 
creep, recovery, recrystallization, 
grain growth, structural phase transformations, and strain-hardening
have already been reported or are currently underway.
The additional, coupled effect of solute diffusion in alloy materials
is naturally incorporated into PFC-type descriptions.
Issues that we believe require further development or should be kept in mind
include choice of ensemble, control of stress-strain-volume relations,
quantification of vacancy concentration, and its connection to climb rates.

\begin{acknowledgments}
This work has been supported by the Natural Science and Engineering 
Research Council of Canada (NSERC), and
supercomputing resources have been provided by CLUMEQ/Compute Canada.
The atomic visualization and analysis packages Ovito \cite{ovito2010}
and the Dislocation Extraction Algorithm (DXA) \cite{dxa2012}
were used in this work.
\end{acknowledgments}


\begin{thebibliography}{77}
\bibitem[{\citenamefont{footnote6}(2014)}]{footnote6}\citenamefont{
Numerical grid spacing $\Delta x=a/(6\sqrt{6})$, time step $\Delta t=0.1$, 
and penalty parameters $L_{\ell}=20\Delta x$, $M_{\ell}=1$, 
$L_P=24\Delta x$, $M_P=0.05$ were also used, where $a=1.8537$.
In Fig.\ \ref{FRdipole}, system dimensions
$N_x=68\sqrt{6}a$, $N_y=117\sqrt{2}a$, and $N_z=37a/\sqrt{3}$ were used
($\sim 2.5\times10^6$ atoms or density peaks),
along with loop dimensions $L_x=10\sqrt{6}a$, $L_z=4\sqrt{3}a$
and shear rate $\dot{\epsilon}_{\rm zy}=0.000235/t$.
In Fig.\ \ref{FRquad}, system dimensions
$N_x=68\sqrt{6}a$, $N_y=80\sqrt{2}a$, and $N_z=55a/\sqrt{3}$ were used
($\sim 2.4\times10^6$ atoms or density peaks),
along with loop dimensions $L_x=5\sqrt{6}a$, $L_z=9\sqrt{3}a$
and shear rate $\dot{\epsilon}_{\rm zy}=0.000235/t$.
}
\bibitem[{\citenamefont{footnote10}(2014)}]{footnote10}\citenamefont{
Numerical grid spacing
$\Delta x=\sqrt{2}a/16$ and time step $\Delta t=0.01$
were used, where $a \simeq 0.817$.
System dimensions were varied to confirm the absence of large
finite size effects, and results are presented for the case of
$N_x=54\sqrt{3}a$, $N_y=38\sqrt{6}a$, and $N_z=66\sqrt{2}a$
($\sim 1.6\times10^6$ atoms or density peaks).
Sphere radii $R_0=2.83a$, $5.66a$, $11.3a$, $16.97a$, $22.63a$, and $28.28a$ 
were examined
with strain rates of $\dot{\epsilon}_{ii}=0.0025/t$ and $0.000125/t$.
The $x$, $y$, and $z$ axes of the simulation cell are aligned with the
$[1\bar{1}1]$, $[\bar{1}12]$, and $[110]$ directions of the BCC crystal,
respectively, and strain was applied separately along each direction. 
A quasi-static case was considered as well, in which the free energy of
a configuration consisting of sphere and dislocation loop with 
radius $R_0$ was measured as a function of
$\epsilon_{ii}$, without any dynamics.
The strain at which the free energy equals that of the undislocated
system is taken to be $\epsilon^*_{ii}$.
}
\bibitem[{\citenamefont{footnote15}(2014)}]{footnote15}\citenamefont{
All parameters are the same as those of Section \ref{subsec:prismaticsources},
except $H_0=0$ (low stacking fault energy, wide dissociation). 
For the $ED$, system dimensions
$N_x=16\sqrt{6}a$, $N_y=87.25\sqrt{2}a$, and $N_z=41\sqrt{3}a$ were used
with $\Delta x=a\sqrt{6}/32$.
For the $SD$, system dimensions
$N_x=28\sqrt{2}a$, $N_y=78\sqrt{6}a$, and $N_z=40\sqrt{3}a$ were used
with $\Delta x=a/(10\sqrt{2})$.
}
\end{thebibliography}
\end{document}